\documentclass[12pt,letterpaper]{article}
\usepackage{amsmath,amssymb}
\usepackage{graphicx,color}

\usepackage[bf]{caption}

\newcommand{\rr}{\mathbb{R}}

\newcommand{\be}{\begin{equation}}
\newcommand{\ee}{\end{equation}}
\newcommand{\ba}{\begin{equationed}}
\newcommand{\ea}{\end{equationed}}
\newcommand{\ben}{\begin{displaymath}}
\newcommand{\een}{\end{displaymath}}
\newcommand{\bea}{\begin{eqnarray}}
\newcommand{\eea}{\end{eqnarray}}
\newcommand{\bean}{\begin{eqnarray*}}
\newcommand{\eean}{\end{eqnarray*}}
\newcommand{\f}{\frac}

\newcommand{\ket}[1]{\left| #1 \right>} 
\newcommand{\bra}[1]{\left< #1 \right|} 

\usepackage{amsmath,amssymb}
\usepackage{graphicx,color}

\def\l {\lambda}

\def\d {\delta}

\definecolor{green}{rgb}{0,0.5,0}

\long\def\symbolfootnote[#1]#2{\begingroup
\def\thefootnote{\fnsymbol{footnote}}\footnote[#1]{#2}\endgroup}

\begin{document}

\begin{titlepage}       \vspace{10pt} \hfill {HU-EP-12/22}

\vspace{20mm}

\begin{center}

{\Large \bf  Bosonic String Quantization in Static Gauge}

\vspace{30pt}

{George Jorjadze,$^{a,\,b}~$
Jan Plefka,$^a$ Jonas Pollok$\,^a$
}
\\[6mm]

{\it\ ${}^a$Institut f\"ur Physik,
Humboldt-Universit\"at zu Berlin,}\\
{\it Newtonstra{\ss}e 15, D-12489 Berlin, Germany}\\[3mm]
{\it${}^b$RMI and Free University of Tbilisi,}\\
{\it Bedia Str.,  Tbilisi, 0183, Georgia}

\vspace{20pt}

\end{center}

\centerline{{\bf{Abstract}}}
\vspace*{5mm}
\noindent
The bosonic string in $D$ dimensional Minkowski space-time is quantized in static gauge.
It is shown that the system can be described by $D-1$ massless free fields constrained on
the surface  $L_m = 0,$  for $m \neq 0 $, where $L_m$ are the generators of conformal transformations.
The free fields are quantized and the physical states are selected by the conditions $L_m|\Psi_{ph}\rangle=0,$ for $m>0$.
The Poincar\'e group generators on the physical Hilbert space are constructed and the
critical dimension $D=26$ is recovered from the commutation relations of the boost operators. The equivalence with the covariant
quantization is established. A possible generalization to the AdS string dynamics is discussed.

\vspace{15pt}
\end{titlepage}

\newpage

\subsection*{Introduction}

In this letter we quantize the bosonic string propagating in flat $D$ dimensional
Minkowski space-time in the static gauge.
 Static gauge relates the target space time coordinate $X^0$ to the evolution parameter $\tau$,
and thus is the most natural gauge of particle or string dynamics.
One would have thought that the subject of this note was already settled in the 1970s and 
should be textbook material by now.
However, to the best of our knowledge the quantization of the string in static gauge
has not been completely achieved to date  
(see \cite{Goddard:1974gd,Marnelius:1975mg,Rohrlich:1974gn,Nikitin:1999zk} for previous literature on the subject).
In a sense the static gauge quantization
is the least optimal route for quantizing the system that one would like to take:
It neither allows for a solution of the constraint conditions, as is achieved in light-cone gauge
(modulo the level-matching condition),  nor is it manifestly covariant, leaving one
with the need to demonstrate quantum Poincar\'e symmetry.

Hamiltonian reduction in this gauge leads to square root expressions for the energy $E$
and the other Poincar\'e symmetry generators
and creates operator ordering ambiguities for them. A possible solution to this problem was proposed in \cite{Dorn:2010wt}
for the particle dynamics in a curved background. In this case, the static gauge naturally leads to
the coordinate representation, where the energy square is quadratic in canonical momenta,
which allows to find a solution of the ordering problem for the operator ${E}^2$ up to a constant factor in front of the scalar curvature term.
The constant can be fixed from the commutation relations of $ E^2$ with other symmetry generators.
The square root from the eigenvalues of ${E}^2$ then provides the energy spectrum. This  quantization scheme in AdS spaces reproduces
the well-known oscillator type spectrum of the AdS particle.

If one tries to apply a similar approach to the string dynamics, a complicated ordering ambiguity problem arises already in a flat background.
Due to this problem, the static gauge was usually avoided in the literature and a consistency of
the quantized string theory was analyzed in the light-cone gauge, where the form of the Poincare group generators is most
simple \cite{Goddard:1973qh} (see \cite{Green:1987sp} for a textbook treatment).

We study quantization of the bosonic string dynamics in static gauge and propose a solution of the ordering problem
similarly to the particle case. For simpicity
we consider open string with worldsheet coordinates $(\tau,\sigma)$ given on the strip $\tau\in \rr^1$, $\sigma\in (0,\pi),$
where the target space coordinates $X^\mu$ $(\mu = 0, \cdots, D-1)$ satisfy the Neumann boundary conditions $X'_\mu(\tau,0)=0=X'_\mu(\tau,\pi).$
The generalization of the obtained results to the closed string dynamics is straightforward.
We use the Minkowski space metric $\eta_{\mu\nu}=\mbox{diag}(-1,1,...,1)$.

\subsection*{Hamiltonian reduction}

The open string dynamics is described by the following action in the first order formulation
\begin{equation}\label{full action}
S = \int \mathrm{d} \tau \int_0^\pi \f{\mathrm{d} \sigma}{\pi} \Big(\mathcal{P}_\mu \dot{X}^\mu - \lambda_1(\mathcal{P}_\mu \mathcal{P}^\mu + X'_\mu X'^\mu) - \lambda_2 (\mathcal{P}_\mu X'^\mu) \Big)~.
\end{equation}
Here $\mathcal{P}_\mu$ are the canonically conjugated variables to the target space coordinates ${X}^\mu$ and the
Lagrange multipliers  $\lambda_1,$  $\lambda_2$ enforce the Virasoro constraints
\begin{equation}\label{Virasoro constraints}
\mathcal{P}_\mu \mathcal{P}^\mu + X'_\mu X'^\mu=0~,\qquad\qquad \mathcal{P}_\mu X'^\mu=0~,
\end{equation}
which are the generators of gauge transformations.

The static (or time-like) gauge  is introduced by the gauge fixing conditions
\begin{equation}\label{static gauge}
X^0 +\mathcal{P}_0 \tau=0~, \qquad\quad \mathcal{P}_0'=0~.
\end{equation}
The action (\ref{full action}) in this gauge reduces to
\begin{equation}\label{reduced action}
S = \int \mathrm{d} \tau \int_0^\pi \f{\mathrm{d} \sigma}{\pi} \left ( \mathcal{P}_k \dot{X}^k -\f{1}{2}\,\mathcal{P}_0^2\right)~,
\end{equation}
where $k=1, \cdots, D-1$ and we have neglected the time derivative term $-\frac{d}{d\tau}\left(\frac{1}{2}\,\mathcal{P}_0^2\tau\right)$.

With the help of \eqref{Virasoro constraints} and \eqref{static gauge}, the term $\mathcal{P}_0^2$ can be expressed through the phase space
variables $(\mathcal{P}_k, {X}^k)$ and we find the free-field Hamiltonian in \eqref{reduced action}
\begin{equation}\label{Hamiltonian}
H=\f{1}{2}\int_0^{\pi}\f{\mathrm{d}\sigma}{\pi}\,\left(\vec{\mathcal{P}}^2 + \vec{X}'^{\,2} \right) ~.
\end{equation}
Thus, the reduced action \eqref{reduced action} describes $D-1$ massless free fields with the constraints
\begin{equation}
\label{reduced Virasoro constraints}
 \left(\vec{\mathcal{P}}^2 + \vec{X}'^{\,2}\right)'= 0~, \qquad \quad \vec{\cal P} \vec{X}' = 0~,
\end{equation}
which still remain from  \eqref{Virasoro constraints} and \eqref{static gauge}. At this stage we stop the Hamiltonian reduction
and analyze the $D-1$ dimensional free-field theory with the constraints \eqref{reduced Virasoro constraints}.

The free fields on the $(\tau,\sigma)$ strip are given by $X^k(\tau,\sigma) = \f{1}{2} \phi^{\,k}(\tau + \sigma) + \f{1}{2}\phi^{\,k}(\tau - \sigma)$,
where the chiral components admit the mode expansion
\begin{equation}
 \phi^{\,k}(z) = q^k + p^k z + \mathrm{i} \sum_{n \neq 0} \f{a_n^k}{n}\,\, \mathrm{e}^{- \mathrm{i} n z} \; , \quad z\in \mathbb{R}\, ,
\end{equation}
with the canonical Poisson brackets
\begin{equation}\label{canonical PB}
\{ p^k, q^l \} = \delta^{kl} \qquad \qquad \{a_m^k, a_n^l \} = \mathrm{i} m\, \delta^{kl} \, \delta_{m+n} ~.
\end{equation}

The generators of conformal transformations defined by
\begin{equation}\label{conformal generators}
L_m = \f{1}{2}\int_0^{2 \pi} \f{\mathrm{d} z}{2 \pi} \,\, \mathrm{e}^{\mathrm{i}mz}\,\vec\phi'(z)^2  = \f{1}{2} \sum_{n = -\infty}^\infty\vec{a}_{m-n} \vec{a}_n \; ,\quad\quad
(\vec{a}_0 = \vec{p})
\end{equation}
realize the Witt algebra
\begin{equation}\label{Witt algebra}
 \{ L_m, L_n \} = \mathrm{i} \, (m-n) \, L_{m+n}~,
\end{equation}
where $L_0$  coincides with the Hamiltonian \eqref{Hamiltonian}.

The constraints (\ref{reduced Virasoro constraints}) may then be seen to be equivalent to
\begin{equation}\label{constraints}
 L_m = 0~,   \qquad \mbox{for} \qquad m\neq0 ~,
\end{equation}
and according to \eqref{Witt algebra} they form a set of second class constraints.

The quantum theory of this system can be identified with a quantized $D-1$ component free-field theory
restricted on the physical states, which satisfy the conditions
\begin{equation}
  \label{physical conditions}
  L_m |{\Psi}_{ph}\rangle = 0~,  \qquad \mbox{for} \qquad m > 0 ~.
\end{equation}

We follow this scheme in the next section, but before quantization let us discuss the realization of the Poincare symmetry in static gauge.
To find the free-field form of the symmetry generators, one can start with the dynamical integrals of the initial system \eqref{full action}
\begin{equation}\label{dynamical integrals}
P^\mu = \int_0^{\pi} \f{\mathrm{d} \sigma}{\pi} \; \mathcal{P}^\mu ~, \qquad
J^{\mu\nu} = \int_0^{\pi} \f{\mathrm{d} \sigma}{ \pi} \; \left( \mathcal{P}^\mu \,X^\nu- \mathcal{P}^\nu \,X^\mu \right)~
\end{equation}
and calculate them in the gauge \eqref{static gauge}, for example, at $\tau=0$. The time component $X^0$ then vanishes and
for $\mathcal{P}^0$ only its zero mode $p^0$ survives, which can be expressed through the free-field variables from the first Virasoro constraint in \eqref{Virasoro constraints}.
This simplification degenerates the boosts $J^{0k}$ and one has to deform them by the constraints
\eqref{constraints}, in order to keep the constraint surface Lorentz invariant.\footnote{Translation and rotation generators do not need any deformations.}
Taking a linear deformation only, we find
\bea\label{dynamical integrals 1}
P^k = p^k ~,&       &J^{kl}= p^k q^l - p^k q^l + \mathrm{i} \sum_{n \neq 0} \f{a_{-n}^k a_n^l}{n}~,\\ \label{dynamical integrals 2}
P^0=p^0=\sqrt{2 L_0}~,&  \quad &J^{0k} =   p^0  q^k+\frac{\mathrm{i}}{p^0 } \sum_{n\neq 0} \f{a_n^k}{n} \,\,L_{-n}~.
\eea

The Poisson brackets of these functions form the Poincare algebra on the phase of the free-field theory and the invariance of the constraint surface
\eqref{constraints} is given by
\begin{equation}\label{PB with L_n}
\{J^{0k},\,L_m\}=
\frac{m}{p^0 }\sum_{n\neq -m,\,\, 0}\frac{a^k_{n+m}}{n+m}\,\,L_{-n}-\left(\frac{\mathrm{i}(mq^k+p^k)}{p^0 }+
\frac{m}{(p^0)^3 } \sum_{n\neq 0} \f{a_n^k}{n} \,\,L_{-n}\right)L_m~.
\end{equation}

Note that the r.h.s. of this equation contains $L_n$'s with both positive and negative indices.
The quantum version of \eqref{PB with L_n}, therefore, can not provide the Lorentz invariance of the physical Hilbert space defined by
\eqref{physical conditions}. Hence, one needs further deformation of the boosts
in terms of higher powers of the constraints
${J}^{0k} \mapsto\mathcal{J}^{0k}={J}^{0k}+C_{n_1,n_2}^k\,L_{-n_1}L_{-n_2}+...~$, to get a suitable form
of the invariance conditions.

Let us look for the deformed boosts in the form
\be\label{boosts}
 \mathcal{J}^{0k} = J^{0k} + \frac{\mathrm{i}}{p^0 } \sum_{j\geq 2}\left(\sum_{n_1,...,n_j}f_j(p^0)\,\,\frac{a_n^k}{n} \,\,L_{-n_1}...L_{-n_j}\right)~,
\ee
where $f_j(p^0)$ are $p^0$ dependent coefficients and $n=\sum_{i=1}^j{n_i}$, with $\,n_i\neq 0$ and $\,n\neq 0$.
The first part of the r.h.s of \eqref{PB with L_n} contains terms which are linear in the constraints
and do not contain $L_m$.  One can cancel these terms in $\{\mathcal{J}^{0k},\,L_m\}$, if  $f_2=\f{1}{2(p^0)^2}$.
This quadratic deformation of $J^{0k}$ changes the factor in front of $L_m$ in the r.h.s. of \eqref{PB with L_n}
and creates new quadratic terms in $\{\mathcal{J}^{0k},\,L_m\}$.
To cancel the new quadratic terms, one can choose the coefficient $f_3$ in \eqref{boosts} and then
continue this process step by step. It leads to the recurrence relations
\be\label{rec relations}
(j+1)(p^0)^2\,f_{j+1}+(2j-1)\,f_j=0~,
\ee
and with $f_1 = 1$, we find
\begin{equation}
 f_{j} = (-)^{j-1}\f{\mathcal{C}_{j-1}}{e^{j-1}}  ~,
\end{equation}
where $e=2(p^0)^2$ and $\mathcal{C}_j=\f{(2j)!}{j! (j+1)!}$ are the Catalan numbers.

In this way we end up with the Poisson brackets of the form
\begin{equation}\label{PB with L_n 1}
\{\mathcal{J}^{0k},\,L_m\}= {\cal{A}}_m^k\,L_m~,
\end{equation}
which may realize the quantum Lorence symmetry on the physical Hilbert space. Note the lack of summation on the r.h.s. of the equation.

\subsection*{Quantization}

We quantize the system by lifting the canonical Poisson brackets \eqref{canonical PB} to the commutation relations
\begin{equation}
      [p^k, q^l] = - \mathrm {i} \, \delta^{kl} \quad \quad [a_m^k, a_n^l] = m \, \delta_{m+n} \, \delta^{kl} \; .
\end{equation}
The unconstrained Hilbert space $\mathcal{H}$ is generated by the action of the creation operators $a^k_{-n}$ onto the
momentum dependent ground state $|\vec p\,\rangle$, which is defined by
\begin{equation}\label{graund state}
a_0^k|\vec p\,\rangle = p^k \ket{\vec p\,} ~,  \quad \quad a^k_{n} \ket{\vec p\,} = 0~,  \qquad  n > 0~.
\end{equation}
The operators $L_n$ have no ordering ambiguity, except $L_0$, and choosing $L_0=\frac{1}{2}\,\vec{p}^{\,2}+N$, where $N$ is the level operator,
\be\label{L_0}
N=\sum_{n>0} \vec{a}_{-n}\,\vec{a}_n~,
\ee
one gets the Virasoro algebra in the standard form
\begin{equation}
  [L_m, L_n]  = (m-n) L_{m+n} + \f{D-1}{12} (m^3-m) \delta_{m+n} ~.
\end{equation}

Let us now describe the physical states defined by \eqref{physical conditions}. Since the commutators of
$L_1$ and $L_2$ generate all other $L_n$'s with positive $n$, it suffices to check only the two conditions
\be\label{L1=0=L2}
L_1|\Psi\rangle=0~,   \qquad        L_2|\Psi\rangle=0~,
\ee
to verify whether a state $|\Psi\rangle$ is physical or not. The ground state $|\vec p\,\rangle$ is obviously physical.
If we write the first level states in the form $|\Psi\rangle=\l^k\,a_{-1}^k|\vec p\,\rangle$, where $\l^k$ are constants, from \eqref{L1=0=L2}
we obtain $\l^k\,p^k=0$. Thus, there are $D-2$ independent components on the first exited level, that is in accordance with other quantization schemes.
The conditions \eqref{L1=0=L2} for the second level physical states 
$|\Psi\rangle=\left(\Lambda^{kl}\,a_{-1}^k\,a_{-1}^l+\rho^k\,a_{-2}^k\right)|\vec p\,\rangle$ lead to $\Lambda^{kl}p^l+\rho^k=0,$  
$\Lambda^{kk}+2\rho^k\,p^k=0$, and the number of independent components here is $\frac{(D-2)(D+1)}{2}$. This is again consistent with the light-cone or covariant quantization
schemes. 

One can continue the description of higher level physical states as in the covariant quantization \cite{Green:1987sp}.
The difference here is that we do not have negative or zero norm states and no factorization over the zero norm states exists. 
A physical state with momentum $\vec p$ and level $N$ will be denoted by $|\vec p,\,N\rangle$.

Let us consider the Poincare symmetry generators. The translation and rotation generators have no ordering ambiguity and one 
can use their classical expressions \eqref{dynamical integrals 1} directly. The ordering freedom for the energy square operator can be expressed by
a parameter $a$ in the form $(p^0)^2=2(L_0-a)$ and we get the energy operator
\be\label{energy operator}
p^0=\sqrt{\vec{p}^{\,\,2} + 2(N-a)}~,
\ee
which is diagonal on the states $|\vec p,\,N\rangle$. The mass operator then becomes $M^2=2(N-a)$.

Most problematic are the boosts. Using their classical expression \eqref{boosts}, the action of the boost operators on a physical state $\ket{\vec p,\,N}$ we represent in the following form
\be\label{boost operators}
 \mathcal{J}^{0 k} \ket{\vec p,\,N} = \left(\colon p^0 q^k \colon + \frac{\mathrm{i}}{p^0} \sum_{n=1}^{N}
\sum_{(n_1,...,n_j)} f^{(n_1,...,n_j)}(p^0)\,\,L_{-n_1} \cdots L_{-n_j} \f{a_n^k}{n}  \right )\ket{\vec p,\,N} ~,
\ee
where $\colon p^0 q^k\colon\equiv\f{1}{2}\left(q^k p^0  +p^0 q^k\right)=q^k p^0 -  \f{\mathrm{i} \,p^k}{2p^0}\,\,$;
$(n_1,...,n_j)$ are the ordered partitions of an integer $n$, with $n_1 + \cdots + n_j=n,\,$  $n \geq n_1 \geq \cdots \geq n_j > 0$ and
the coefficients $f^{(n_1,...,n_j)}$ are the quantum counterparts of $f_j$.

We allow quantum deformations for these coefficients
and to find them we require the Lorentz invariance of the physical Hilbert space. This implies that the states  $ \mathcal{J}^{0 k} \ket{\vec p,\,N}$
are physical, i.e. they satisfy the conditions \eqref{L1=0=L2}, and the Lorentz algebra
\be\label{[J,J]}
[\mathcal{J}^{0k}, \mathcal{J}^{0l}]\ket{\vec p,\,N}=\mathrm{i} \, J^{kl}\ket{\vec p,\,N}
\ee
is fulfilled. Since the operators $\mathcal{J}^{0 k}$ preserve the level $N$, eq. \eqref{[J,J]} is equivalent
to the relation $\bra{N}[\mathcal{J}^{0k}, \mathcal{J}^{0l}]\ket{\vec p,\,N}=\bra{N}\mathrm{i} \, J^{kl}\ket{\vec p,\,N}$,
where  $\bra{N}$ is a `bra' physical state of the level $N$.
Note that the calculation of these matrix elements can be simplified due to
\be\label{<JJ>}
\bra{N} \mathcal{J}^{0k} \mathcal{J}^{0l} \ket{\vec p,\,N} = \bra{N} \colon p^0 q^k \colon \mathcal{J}^{0l} \ket{\vec p,\,N} \; ,
\ee
which follows from the structure of the boost operators in \eqref{boost operators}.
Using these conditions, one can fix the coefficients $f^{(n_1,...,n_j)}$ level by level.

The Lorentz invariance of the vacuum state is obvious. On the first excited level we have
\bea\label{boost on level 1}
   \mathcal{J}^{0 k} \ket{\vec p,\,1} =
  \left( q^k p^0 -  \f{\mathrm{i} \,p^k}{2p^0} +
\f{\mathrm{i}\,f^{(1)}}{p^0}\, L_{-1} a^k_1  \right )
\ket{\vec p,\,1} \; ,
\eea
where $p^0=\sqrt{\vec{p}^{\,\,2} + 2(1-a)}$. In this case one has to check only the first condition \eqref{L1=0=L2}
\be\label{1 level invariance}
 L_1 \mathcal{J}^{0k} \ket{\vec p,\,1} = [L_1,\mathcal{J}^{0k}] \ket{\vec p,\,1} = 0 \; ,
\ee
and using that $[L_1,q^k]=-\mathrm{i}a^k_1$, we find
\be\label{coeff 1}
 f^{(1)} = \f{\vec{p}^{\,\,2} + 2(1-a)}{\vec{p}^{\,\,2}} ~.
\ee
Due to \eqref{<JJ>}, the check of \eqref{[J,J]} is reduced to
\be\label{1 level check}
\bra{\,1}  p^0 q^k \,\,  \f{\mathrm{i}\,f^{(1)}}{p^0}\,  L_{-1} a^l_1  \ket{\vec p,\,1} = -\bra{\,1} a_{-1}^k a^l_1  \ket{\vec p,\,1} \; ,
\ee
which provides $f^{(1)} = 1$ and at the same time fixes the ordering constant $a=1$.
The corresponding mass operator $M^2 = 2(N - 1)$  then reproduces the bosonic string spectrum.

On the second level eq. \eqref{boost operators} takes the form
\be\label{2 level calcul}
  \mathcal{J}^{0k} \ket{\vec p,\,2} = \left(q^k p^0 -  \f{\mathrm{i} \,p^k}{2p^0}
  +\f{ \mathrm{i}}{p^0}\, L_{-1} a^k_1  + \f{\mathrm{i}\,f^{(2)}}{2 \, p^0}\,  L_{-2} a^k_2  +
\f{\mathrm{i}\,f^{(1,1)}}{2 \, p^0}  L_{-1} L_{-1} a^k_2  \right ) \ket{\vec p,\,2} ~,
\ee
with $p^0=\sqrt{\vec{p}^{\,\,2} + 2}$, and the conditions $L_1 \mathcal{J}^{0k} \ket{\vec p,\,2} = 0,$ $~L_2 \mathcal{J}^{0k} \ket{\vec p,\,2} = 0$
are equivalent to
\be\label{2 level conditions}
3 f^{(2)} + 2(\vec{p}^{\,\,2}+1) f^{(1,1)} =2 ~,\quad \qquad
(4\vec{p}^{\,\,2} + D-1) f^{(2)} + 6 \vec{p}^{\,\,2} f^{(1,1)}=4(\vec{p}^{\,\,2}+5) \; .
\ee
The check of \eqref{[J,J]} now is reduced to
\be\label{2 level commut}
\bra{\,2}  p^0 q^k \,  \left(\f{\mathrm{i} \,f^{(2)}}{2 p^0}\,\,L_{-2} a^l_2  +
\f{\mathrm{i} \,f^{(1,1)}}{2 p^0}\,\, L_{-1}L_{-1} a^l_2\right ) \ket{\vec p,\,2} =
-\f{1}{2}\bra{\,2}  a_{-2}^k a^l_2  \ket{\vec p,\,2} \; ,
\ee
and it gives the additional equation $f^{(2)} - f^{(1,1)} = 1$. These three equations define the
coefficients of the second level
\be\label{coefficinets 2}
f^{(1,1)} = -\f{1}{e + 1} ~, \quad \quad f^{(2)} = \f{e}{e + 1} ~,\qquad \mbox{with} \quad e=2(p^0)^2,
\ee
and also fix the critical space-time dimension  $D=26$.

Using the coefficients of the first two levels
and the obtained values of the parameters $a$ and $D$, for the third level coefficients we similarly obtain
\be\label{coefficinets 3}
f^{(1,1,1)} = \f{2}{(e + 1)(e+4)} ~, \quad f^{(2,1)} = -\f{e}{(e + 1)(e + 4)} ~,\quad f^{(3)} = \f{e^2+e}{(e + 1)(e + 4)}~.
\ee
This procedure can be continued step by step and if one knows all coefficients up to the level $N-1$, one can calculate
the coefficients on of the level $N$. This construction provides the following structure
\be\label{coeffficinets f}
  f^{(n_1,...,n_j)} =  \frac{\mathcal{P}_{n-j}^{(n_1,...,n_j)}(e)}{\displaystyle
  \prod_{m=1}^{n-1} (e+m^2)}~,
\ee
where $\mathcal{P}_{n-j}^{(n_1,...,n_j)}(e)$ is a polynomial of the degree $n-j$ with a partition dependent coefficient.
To prove \eqref{coeffficinets f}, one can project the `ket' state $\mathcal{J}^{0 k} \ket{\vec p,\,N}$ onto a `bra' state of the level $N$
$\langle\,L_{-m_1} \cdots L_{-m_{i}}\,0|,$ defined by a partition $(m_1,...,m_i)$.
This projection has to vanishes, since the state $\mathcal{J}^{0 k} \ket{\vec p,\,N}$ has to be physical. Hence, the coefficients of the level $N$
satisfy the equations
\be\label{eq for f}
\mathcal{M}_{(m_1,...,m_i)(n_1,...,n_j)}f^{(n_1,...,n_j)}=F_{(m_1,...,m_i)}~.
\ee
Here $\mathcal{M}$ is the matrix with the coefficients $\bra{0}L_{m_i} \cdots L_{m_1} L_{-n_1} \cdots L_{-n_j}\ket{0}$
and $F_{(m_1,...,m_i)}$ is obtained from the lower level coefficients. The determinant of the matrix $\mathcal{M}$ is just
the Kac determinant, and its simple form \cite{Ginsparg:1988ui} for the central charge $c=D-1=25$ provides  the denominator 
in \eqref{coeffficinets f}. 

We could not find a closed form of the numerator for an arbitrary partition, 
though in some cases they are calculable exactly by recurrence relations.
For example, in case of $f^{(1, \cdots,1)}$ the Polynomial in the numerator of \eqref{coeffficinets f} degenerates to a constant and, as in the classical case,
it is given by the Catalan number
\be\label{coeff 1,1,1}
\mathcal{P}_{0}^{(1,...,1)} = (-1)^{N-1} {C_{N-1}} ~.
\ee
Indeed, we explicitly computed the $f^{(n_1,...,n_j)}$ coefficients up to the level 8.

\subsection*{Connection to the covariant quantization}

The covariant quantization \cite{Green:1987sp} contains additional `oscillator' degrees of freedom 
created by the time-component operators $a_{-n}^0$, with $n>0$. The norm of the corresponding states is indefinite
due to the commutation relations $[a_m^0,a_n^0]=-m\d_{m+n}$. 

We use the notation $||\,\cdot\,\rangle\rangle$  for the `bra' states of the covariant quantization, to distinguish them
from the static gauge states. The physical states of the covariant quantization satisfy the conditions
\be\label{covariant psi}
\hat L_m||{\psi}_{ph}\rangle\rangle=0~,  \quad \mbox{for} \quad m>0, \qquad  \mbox{and} \quad (\hat L_0-1)||{\psi}_{ph}\rangle\rangle=0~,
\ee
where
\be\label{covariant L_n}
\hat L_m=L_m-L_m^0~, \qquad \mbox{with} \qquad  L_m^0=\frac{1}{2}\,a_{m-n}^0\,a_n^0~.
\ee
The physical states of the static gauge quantization, therefore, are the physical states of
the covariant quantization with non excited time-component degrees of freedom.
 In particular, the state $\ket{\vec p,\,N}$ can be identified with the states $||{p^0,\vec p\,;\,0,N}\rangle\rangle$, where $0$ denotes
 the vacuum state in the time-component sector and $p^0$ is given by \eqref{energy operator} at $a=1$.
 
The boost operators in the covariant quantization 
\be\label{covarian boost}
J^{0k}=p^0q^k-p^kq^0+\mathrm{i}\sum_{n> 0}\left(\f{a_{-n}^0a_n^k}{n}-\f{a_{-n}^ka_n^0}{n}\right)
\ee 
have no ordering ambiguity and their action on the physical state $||{p^0,\vec p\,;\,0,N}
\rangle\rangle$ is given by a finite sum in \eqref{covarian boost}
with $n\leq N$.
Let us take the $n$-th term of this sum, change the ordering in  $a_{-n}^0\,a_n^k$ and consider
the action of $a_{-n}^0$ on $||{p^0,\vec p\,;\,0,N}\rangle\rangle$. The operator $a_{-n}^0$ then creates an $n$-level state in the time-component sector and
one can write the expansion
\be\label{expansion}
a_{-n}^0||{p^0,\vec p\,;\,0,N}\rangle\rangle=\sum_{(n_1,...,n_j)} \tilde f^{(n_1,...,n_j)}(p^0)\,\,L_{-n_j}^0 \cdots L_{-n_1}^0\,||{p^0,\vec p\,;\,0,N}\rangle\rangle~,
\ee
where $\tilde f^{(n_1,...,n_j)}(p^0)$ are the expansion coefficients and $(n_1,...,n_j)$ is an ordered partitions of $n$ as in \eqref{boost operators}.
However, note that we now chose the opposite ordering.  The action of the operator $L_{-n_1}^0$ on the physical
state $||{p^0,\vec p\,;\,0,N}\rangle\rangle$ can be replaced by $L_{-n_1}$, and since $L_{-n_1}$  commutes with the $L_{-n}^0$ operators, 
we can move $L_{-n_1}$ to the left in \eqref{expansion}. With this replacement procedure in \eqref{expansion}, 
the action of the operator \eqref{covarian boost} onto the
physical state $||{p^0,\vec p\,;\,0,N}\rangle\rangle$ takes the structure \eqref{boost operators},  and comparing these two expressions we conclude that 
 $f^{(n_1,...,n_j)}(p^0)=p^0 \tilde f^{(n_1,...,n_j)}(p^0)$. This shows the equivalence between the two quantizations.
 
To our knowledge, the expansion coefficients in  \eqref{expansion} are not known in a closed form,  even for the space-component excited states.  

\subsection*{Conclusion}

We performed a quantization of the bosonic strings in static gauge. 
It was shown that the string dynamics in $D$ dimensional Minkowski space can be described by $D-1$ component conformal free-field theory,
restricted on the constraint surface \eqref{constraints}. 

Most problematic in this approach are the boost operators.
Their structure has been found on the basis of classical calculations. This structure defines the boosts up to some energy dependent coefficients.
These coefficients can be calculated to any desirable level, but their closed form is still missing.
Low level calculations of the commutation relations of the boost operators provide the string mass spectrum and the space-time critical dimension
in a very simple manner. 

We have obtained the equivalence between the static gauge and the covariant quantization. 
This equivalence shows that the coefficients we were looking for in the static gauge quantization 
are just the expansion coefficients of the oscillator excitations in terms of the Virasoro excitations.
A solution of this problem could be useful for the calculation of the Liouville $S$-matrix \cite{Zamolodchikov:1995aa}, 
where the Virasoro generators are deformed by the terms linear in the creation-annihilation operators .

The initial motivation of this work was a continuation of the paper \cite{Dorn:2010wt} and the investigation of the AdS string in static gauge.
Even though the static gauge is not a conformal gauge for AdS strings,  
the Hamiltonian reduction can be done similarly to the flat background. The reduced picture here exhibits a new coset WZW structure, which differs
from the Pohlmeyer reduction \cite{Grigoriev:2007bu} and we plan to investigate it in the future.

\vspace{5mm}

\noindent
{\bf Acknowledgments}

\vspace{3mm}

\noindent
We thank Harald Dorn, Ruslan Metsaev and Vladimir Mitev for useful discussions.
This work was supported by the VolkswagenStiftung.


\begin{thebibliography}{99}
\addtolength{\parskip}{-1ex}


\bibitem{Goddard:1974gd}
  P.~Goddard, A.~J.~Hanson and G.~Ponzano,
 \emph{ ``The Quantization of a Massless Relativistic String in a Timelike Gauge,''}
  Nucl.\ Phys.\ B {\bf 89} (1975) 76.
  

\bibitem{Marnelius:1975mg}
  R.~Marnelius,
 \emph{ ``Treatment of the Classical Relativistic String in Any Orthonormal Gauge,''}
  Nucl.\ Phys.\ B {\bf 104} (1976) 477.


\bibitem{Rohrlich:1974gn}
  F.~Rohrlich,
 \emph{ ``Quantum Dynamics of the Relativistic String,''}
  Phys.\ Rev.\ Lett.\  {\bf 34} (1975) 842.


\bibitem{Nikitin:1999zk}
  I.~Nikitin,
 \emph{ ``String theory in Lorentz invariant timelike gauge,''}
  hep-th/9907196.
  
\bibitem{Dorn:2010wt}
  H.~Dorn, G.~Jorjadze, C.~Kalousios and J.~Plefka,
 \emph{ ``Coordinate representation of particle dynamics in AdS and in generic static spacetimes,''}
  J.\ Phys.\ A A {\bf 44}, 095402 (2011)
  [arXiv:1011.3416 [hep-th]].




\bibitem{Goddard:1973qh}
  P.~Goddard, J.~Goldstone, C.~Rebbi and C.~B.~Thorn,
\emph{  ``Quantum dynamics of a massless relativistic string,''}
  Nucl.\ Phys.\ B {\bf 56} (1973) 109.

\bibitem{Green:1987sp}
  M.~B.~Green, J.~H.~Schwarz and E.~Witten,
 \emph{ ``Superstring Theory. Vol. 1: Introduction,''}
  Cambridge, Uk: Univ. Pr. ( 1987) 469 P. ( Cambridge Monographs On Mathematical Physics)

\bibitem{Ginsparg:1988ui}
  P.~H.~Ginsparg,
 \emph{ ``Applied Conformal Field Theory,''}
  hep-th/9108028.

\bibitem{Zamolodchikov:1995aa}
  A.~B.~Zamolodchikov and A.~B.~Zamolodchikov,
\emph{``Structure constants and conformal bootstrap in Liouville field theory,''}
  Nucl.\ Phys.\ B {\bf 477} (1996) 577
  [hep-th/9506136].

  
\bibitem{Grigoriev:2007bu}
  M.~Grigoriev and A.~A.~Tseytlin,
\emph{  ``Pohlmeyer reduction of $\mbox{AdS}_5 \times S^5$ superstring sigma model,''}
  Nucl.\ Phys.\  B {\bf 800} (2008) 450
  [arXiv:0711.0155 [hep-th]].



\end{thebibliography}
\end{document}